\documentclass[12pt,twocolumn]{article}
\begin{document}

\title{Locator/identifier split using the data link layer}

\author{Victor Grishchenko, Ural State University}

\maketitle

\section{Introduction}

The locator/identifier split approach assumes separating functions of a locator (i.e. topology--dependent attachment point address) and identifier (topology-independent unique identifier) currently both served by an IP address. This work is an attempt to redefine semantics of MAC address to make it a pure layer-2 locator instead of a pure globally-unique identifier. Such an exercise might be interesting from the standpoint of Ethernet scaling and Metro Ethernet technologies. From the global routing perspective, introduction of multihoming, traffic engineering and failover at the 2nd layer may reduce pressure on the 3rd layer.

Historically, an Ethernet network was supposed to be a single wire with many devices attached. For reliable identification, each device has a factory-preset globally-unique 6-byte MAC address. Each Ethernet frame has destination and source MAC addresses. Currently, switched networks are prevalent, so one physical ``wire'' normally connects just two devices. Layer-2 switched networks are typically divided into flat logical segments (VLANs) where MAC addresses are used as pure identifiers to perform frame forwarding/routing using spanning trees and announce flooding. Interestingly, that behavior is more akin to what was traditionally considered as ``routing'', albeit ultimately it still emulates shared copper.

The trick is to try to introduce advanced routing options and topologies to the data link layer of a given network, without touching anything at the 3rd layer or end hosts. 

\section{Redefining MAC}

The objective is to ease layer-2 switching in large layer-2 networks by overloading MAC addresses to be pure locators. IP address thus plays as a pure identifier.

As a matter of fact, mesh networks are not effective. STP protocol, for example, starts with deactivating extra links to turn the topology into  a tree. I will consider a tiered architecture where every switch is connected to some uplink switches and some downlink switches/devices. ``Horizontally'' connected switches are modeled as a single switch (``stacking''). Shortcutting, a weaker form of ``horizontal'' linking, is modeled as a fictive common uplink switch. 

The proposed addressing scheme is a feature-cut of the prefix-bunch architecture. A device of $i$-th tier has a number of ``BigMAC'' addresses consisting of $i$ meaningful bytes and $6-i$ padding zeroes: $\{b_{1}{\ldots}b_{i}~0\ldots\}$. More precisely, each such address belongs to some uplink port. Further, $c$-th downlink port is assigned addresses $\{b_{1}{\ldots}b_{i}~c~0 \ldots\}$. 

Thus, every BigMAC address corresponds to a downward path from some top-tier switch to the target device. Differently from hierarchical addressing, all the network's addresses combined form not a tree, but a tree-resembling structure I will christen a ``branch bunch''. The average number of addresses an end host will have is estimated as $N^{\frac{\log_{2}u}{\log_{2}d-\log_{2}u}}$, where $N$ is the number of end hosts; $u$ and $d$ are uplink/downlink fanouts resp. (So, $\sqrt[4]{N}$ for $u=2$, $d=32$.) It is not generally supposed that a device knows all of its addresses.

\section{Switching}

Obviously, forwarding a frame from uplink to downlink is as simple as checking $(i+1)$-th byte of the destination address which contains the number of the egress port. To forward a frame from a downlink up, a switch has to direct the frame to the uplink port that owns the source BigMAC address (strategy $\alpha$). 
The switch might as well rewrite first $i$ bytes of the source address to forward the frame to an arbitrary uplink -- assuming, that the top-tier switches are fully interconnected (strategy $\beta$).
A more sophisticated upward-forwarding strategy is to check the destination address against available uplink addresses to detect the longest prefix match (strategy $\gamma$). Albeit, this functionality is not dirt-cheap to implement, so it is better to shift it away from the switch, see Sec.~\ref{sec:arpdhcp}.

Another strategy $\delta$ may reduce upward forwarding to the same kind of byte-check employed by downward forwarding. Namely, if a typical switch has 24\ldots32 ports and uses a byte of addressing space, then there are 3 spare bits to use. As the number of uplinks is supposedly less than 8, 3 bits of the byte may stand for the number of the uplink the original BigMAC came from, while the rest 5 denote the downlink port, as before. So, 3 bits of the $(i+1)$-th byte of the source address denote the egress uplink port. 

One more issue is when to forward from downlink to downlink. One criterion is first $i$ bytes of the source address being equal to the first $i$ bytes of the destination address.

To preserve compatibility with the end hosts, some MAC address rewriting is needed. By using ARP, end hosts learn BigMACs of peers (see Sec.~\ref{sec:arpdhcp}). To fully control host-to-host traffic paths, we have to set both source and destination BigMACs. So, a customer-edge switch has not only to rewrite genuine MAC of an end host for a BigMAC, but also to remember which particular source BigMAC to use for a given destination BigMAC. By sacrificing one byte of BigMAC this is also reduced to a byte-check (left as an excercise).

\section{ARP\&DHCP} \label{sec:arpdhcp}

To remain backward compatible with Ethernet+IP end host stacks, a different functioning of ARP and DHCP is needed. I suppose that all ARP and DHCP traffic is diverted to some dedicated ARP\&DHCP server. Possible variants of distributed/tiered implementation are omitted. 

All ARP/DHCP broadcasts of end hosts are upward-flooded, i.e. sent to every uplink. Finally, ARP\&DHCP server gets a copy of a request from every possible path, thus passively learning the topology. The total amount of requests is thus $O(N^{1.25})$ for the reference case of $u=2, d=32$. A reply travels by a single path to the end host.

As mentioned, ARP server may do some traffic engineering by sending a reply containing BigMAC of the target host to a particular BigMAC address of the requesting host so the edge switch learns the association. This way we may ``outsource'' the aforementioned longest-prefix matches to a dedicated out-of-band entity and cache them later on (benefits of $\gamma$ for the price of $\delta$). That also opens possibilities for load balancing.

Some failover and on-the-fly reconfiguration functionality might be achieved by the means of ARP announcements.

\section{Conclusion}

So, if nothing important was overlooked,  branch-bunch locators may bring many gains to the data link layer. Switching logic is dramatically simplified; it needs no routing tables, no associative memory lookups, no longest prefix matches. Scalability is high as forwarding-related computational load on a single switch generally does not depend on the size of the network. The network has simple tools for basic traffic engineering: on-the-fly load balancing and failover. Last but not least, the approach preserves backward compatibility.
Anyway, any questions, comments, criticisms and considerations are welcomed.

\end{document}